\pdfoutput=1
\documentclass[aps,prb,superscriptaddress,twocolumn]{revtex4-1}

\usepackage[colorlinks=true, pdfstartview=FitV, linkcolor=blue, citecolor=blue, urlcolor=blue]{hyperref}
\usepackage[dvipdfmx]{graphicx}
\usepackage{amsmath,amsthm}
\usepackage{color}
\graphicspath{{./figure/}}

\newcommand{\cT}{{\cal T}}

\newcommand{\up}{\uparrow}
\newcommand{\down}{\downarrow}
\newcommand{\be}{\begin{equation}}      
\newcommand{\ee}{\end{equation}}      
\newcommand{\bea}{\begin{eqnarray}}      
\newcommand{\eea}{\end{eqnarray}}

\newcommand{\pf}{\mathrm{pf}}

\newcommand{\Tr}{\mathrm{Tr}}

\begin{document}

\title{Quantum Monte Carlo simulation of a two-dimensional Majorana lattice model}

\author{Tomoya Hayata}
\affiliation{Department of Physics, Chuo University, Tokyo, 112-8551, Japan}
\author{Arata~Yamamoto}
\affiliation{Department of Physics, The University of Tokyo, Tokyo 113-0033, Japan}

\date{\today}

\begin{abstract}
We study interacting Majorana fermions in two dimensions as a low-energy effective model of a vortex lattice in two-dimensional time-reversal-invariant topological superconductors.
For that purpose, we implement ab-initio quantum Monte Carlo simulation to the Majorana fermion system in which the path-integral measure is given by a semi-positive Pfaffian.
We discuss spontaneous breaking of time-reversal symmetry at finite temperature.
\end{abstract}

\maketitle

\section{Introduction}
Realization of Majorana fermions, which are exotic fermions that behave as their own anti-particles (holes), 
has been intensively discussed in condensed matter systems such as fractional quantum Hall (Moore-Read) states~\cite{MOORE1991362}, 
topological superconductors (superfluids)~\cite{PhysRevB.61.10267,PhysRevLett.86.268,PhysRevB.78.195125,PhysRevLett.102.187001}, 
and surface states of topological insulators under the proximity effect between trivial superconductors~\cite{PhysRevLett.100.096407} (For reviews, see e.g., Refs.~\cite{RevModPhys.82.3045,RevModPhys.83.1057,JPSJ.85.022001}).
In these materials, Majorana fermions emerge as collective excitations localized at surfaces or defects (quantum vortices).

Majorana zero modes or Majorana bound states localized at quantum vortices cost zero energy, 
and lead to large ground-state degeneracy~\cite{MOORE1991362,PhysRevB.61.10267,PhysRevLett.86.268}.
The degenerate ground-state wave function has the non-abelian Berry phase.
When the position of two vortices are adiabatically rearranged, it results in the so-called braiding statistics~\cite{NAYAK1996529}, which is considered to be potentially useful for quantum computation~\cite{Kitaev20032,1063-7869-44-10S-S29,RevModPhys.80.1083}.

As the number of vortices increases, the vortices form lattice structure such as the Abrikosov lattice.
Low-energy dynamics of such a vortex lattice in topological superconductors is effectively described as a many-body system of Majorana zero modes.
The system obeys the lattice Hamiltonian describing the inter-vortex tunnelings and interactions of them, which is similar to the Hubbard Hamiltonian.
Revealing the ground-state property of such a many-body Majorana system is important not only for applications of the vortex lattices to quantum computation
but also for understanding new phases of strongly correlated electron systems where the fundamental degrees of freedom are Majorana zero modes.

However, previous studies have mostly focused on one-dimensional Majorana chains, which can be solved by field-theoretical techniques such as bosonization, or by numerical techniques such as exact diagonalization and density matrix renormalization group~\cite{Grover280,PhysRevLett.115.166401,PhysRevB.92.235123,PhysRevB.91.165402,PhysRevB.92.115137}.
It is hard to implement these techniques to higher-dimensional systems, particularly to large two-dimensional systems although two dimensions is often critical dimension in competition between disorder by thermal/quantum fluctuation and order by mean-field dynamics.
Thus ab-initio simulation is required to precisely determine the phase structure.

In this paper we study non-perturbative dynamics of interacting Majorana fermions by utilizing the quantum Monte Carlo simulations which are based on the path-integral formalism of lattice field theories.
We consider an effective Majorana lattice model describing the low-energy behavior of the vortex lattice in topological superconductors with time-reversal symmetry.
By applying the quantum Monte Carlo method to the effective model, we discuss spontaneous breaking of time-reversal symmetry at finite temperature.
Time-reversal symmetry protects Majorana zero modes  against opening a gap, and gapless (free) Majorana fermion states are realized in symmetric phase.
On the other hand in broken phase, double Majorana fermions form a pair and behave as a single spinless Dirac fermion with dynamically generated mass.
We note that the understanding of time-reversal symmetry and its spontaneous breaking is important for condensed matter realization of supersymmetry~\cite{Grover280,PhysRevLett.115.166401,PhysRevB.92.235123}.

The rest of the paper is organized as follows.
In Sec.~\ref{secmodel}, we introduce the Hamiltonian and the path integral of the effective Majorana lattice model.
By using the effective model, we first discuss the phase structure on the basis of the mean-field theory in Sec.~\ref{secMF}, 
and then show the result of quantum Monte Carlo simulation in Sec.~\ref{secQMC}.
Finally we summarize this paper in Sec.~\ref{sec:sum}.

\section{Model}
\label{secmodel}
Let us introduce a low-energy effective model describing quantum vortex lattice in two-dimensional time-reversal-invariant topological superconductors.
A quantum vortex supports a pair of Majorana zero modes $\psi_{\up,\down}$ with spin up and down~\cite{PhysRevLett.102.187001}, which are related by time-reversal symmetry as
\be
\cT^{-1}\psi_\up\cT=\psi_\down,\;\;\cT^{-1}\psi_\down\cT=-\psi_\up.
\ee 
Time-reversal symmetry is anti-unitary operation, namely, $\cT^{-1}i\cT=-i$.

We consider the model Hamiltonian
\begin{equation}
\begin{split}
 \mathcal{H}
=& \sum_\mu it \{ \psi_\uparrow(x) \psi_\uparrow(x+\hat{\mu}) - \psi_\downarrow(x) \psi_\downarrow(x+\hat{\mu}) \}
\\
&- \sum_\mu g \psi_\uparrow(x) \psi_\uparrow(x+\hat{\mu}) \psi_\downarrow(x) \psi_\downarrow(x+\hat{\mu}) 
,
\end{split}
\label{eqH}
\end{equation}
where $\hat{\mu}$ is the unit lattice vector in the $\mu$ direction ($\mu=1$, 2).
The first term describes the inter-vortex hopping and the second term describes the nearest-neighbor interaction between the spin-up and spin-down Majorana fermions.
Both the hopping term and the interaction term keep time-reversal symmetry.
The unique local term $i \psi_\up(x)\psi_\down(x)$, which can open a gap in energy spectrum, is prohibited by time-reversal symmetry since this term is time-reversal odd.
We consider attractive interaction $g>0$, which favors pairing of Majorana fermions.
For the attractive interaction, this model is free from the fermion sign problem for any lattice structure in any dimension, as shown in Sec.~\ref{secQMC}.
We study this model on a two-dimensional square lattice.

A pair of two Majorana fermions can be rewritten by a single spinless Dirac fermion as $\psi(x) = \{\psi_\up(x) +i\psi_\down(x) \}/\sqrt{2}$.
The Hamiltonian \eqref{eqH} is rewritten as
\begin{equation}
\begin{split}
 \mathcal{H}
=& \sum_\mu it \{ \psi(x) \psi(x+\hat{\mu}) + \psi^*(x) \psi^*(x+\hat{\mu}) \}
\\
&- \sum_\mu g \left(n(x)-\frac{1}{2}\right) \left(n(x+\hat{\mu}) -\frac{1}{2}\right)
,
\end{split}
\label{eqHD1}
\end{equation}
where $n(x)=\psi^*(x) \psi(x)$.
For a bipartite lattice such as a square lattice, the Hamiltonian can be further rewritten by the particle-hole transformation, $\psi(x) \to i^{x_1+x_2} \psi(x)$ on even sites $x$ and $\psi(x) \to (-i)^{x_1+x_2} \psi^*(x)$ on odd sites $x$.
The model is equivalent to the spinless repulsive Hubbard model at half-filling\cite{PhysRevB.29.5253,PhysRevB.32.103}
\begin{equation}
\begin{split}
 \mathcal{H}
=& - \sum_\mu t \{ \psi^*(x) \psi(x+\hat{\mu}) + \psi^*(x) \psi(x+\hat{\mu}) \}
\\
&+ \sum_\mu g \left(n(x)-\frac{1}{2}\right) \left(n(x+\hat{\mu}) -\frac{1}{2}\right)
,
\end{split}
\label{eqHD2}
\end{equation}
which was shown to be Majorana positive \cite{2015PhRvB..91x1117L}.
The pairing term $i \psi_\uparrow(x) \psi_\downarrow(x)$ in the original Hamiltonian \eqref{eqH} corresponds to the Dirac mass term $\psi^*(x)\psi(x)-1/2$ in the Hamiltonian \eqref{eqHD1} and the staggered mass term $(-1)^{x_1+x_2} \left(\psi^*(x) \psi(x)-1/2\right)$ in the Hamiltonian \eqref{eqHD2}.

We introduce the Euclidean path integral
\begin{equation}
 Z = \int D\psi_\uparrow D\psi_\downarrow \ e^{-S} ,
\end{equation}
with the Euclidean action
\begin{equation}
 S
= \int d\tau \sum_x \bigg[
\psi_\uparrow \frac{\partial}{\partial\tau} \psi_\uparrow + \psi_\downarrow  \frac{\partial}{\partial\tau} \psi_\downarrow
+ \mathcal{H} \bigg]
.
\end{equation}
The integral of imaginary time $\tau$ is anti-periodic with the period $1/T$.
In the path-integral formalism, the Majorana fermions $\psi_{\up,\down}$ are real Grassmann fields.
In the following sections, we evaluate this path integral by using the mean-field theory and the quantum Monte Carlo simulation.

\section{Mean field}
\label{secMF}

First we study this model in the mean-field approximation.
By the Hubbard-Stratonovich transformation, we obtain
\begin{equation}
 Z = \int D\psi_\uparrow D\psi_\downarrow DC \ e^{ - S' - S_C } ,
\end{equation}
with the bilinear fermion action
\begin{equation}
\begin{split}
S'
=& \int d\tau \sum_x \bigg[
\psi_\uparrow(x) \frac{\partial}{\partial\tau} \psi_\uparrow(x) + \psi_\downarrow(x)  \frac{\partial}{\partial\tau} \psi_\downarrow(x)
\\
&+ \sum_\mu it \{ \psi_\uparrow(x) \psi_\uparrow(x+\hat{\mu}) - \psi_\downarrow(x) \psi_\downarrow(x+\hat{\mu}) \}
\\
&- \sum_\mu iC_\mu(x) \{ \psi_\uparrow(x) \psi_\downarrow(x) + \psi_\uparrow(x+\hat{\mu}) \psi_\downarrow(x+\hat{\mu}) \} \bigg] ,
\end{split}
\end{equation}
and the auxiliary field action
\begin{equation}
S_C = \int d\tau \sum_x \sum_\mu \frac{1}{2g} C_\mu^2 (x)
.
\end{equation}
Assuming that the mean field is homogeneous and isotropic, $C_\mu(x) = C$, we obtain the gap equation
\begin{equation}
\frac{1}{2g} C 
= \langle i \psi_\uparrow \psi_\downarrow \rangle  
= \Tr \left( \frac{\sigma_2}{2} \frac{1}{M} \right)
.
\label{eqgap}
\end{equation}
The fermion matrix $M$ is defined by the fermion action in the matrix notation
\begin{equation}
\begin{split}
 S' &\equiv \frac{1}{2} \Psi^\top M \Psi
\\
&= \frac{1}{2}
\begin{pmatrix}
\psi_\uparrow^\top & \psi_\downarrow^\top
\end{pmatrix}
\begin{pmatrix}
M_{\uparrow\uparrow} & M_{\uparrow\downarrow}
 \\
M_{\downarrow\uparrow} & M_{\downarrow\downarrow}
\end{pmatrix}
\begin{pmatrix}
\psi_\uparrow
\\
\psi_\downarrow
\end{pmatrix} ,
\end{split}
\end{equation}
with the matrix elements
\begin{eqnarray}
 M_{\uparrow\uparrow} &=& 2 \frac{\partial}{\partial\tau} + \sum_\mu  it ( P_{+\mu} - P_{-\mu} ) ,
\\
 M_{\downarrow\downarrow} &=& 2 \frac{\partial}{\partial\tau} - \sum_\mu it ( P_{+\mu} - P_{-\mu} ) ,
\\
 M_{\uparrow\downarrow} &=& - M_{\downarrow\uparrow} = - i4 C ,
\end{eqnarray}
and $P_{\pm\mu} \psi_{\up,\down}(x) = \psi_{\up,\down}(x\pm\hat{\mu})$.

\begin{figure}[t]
 \includegraphics[width=.48\textwidth]{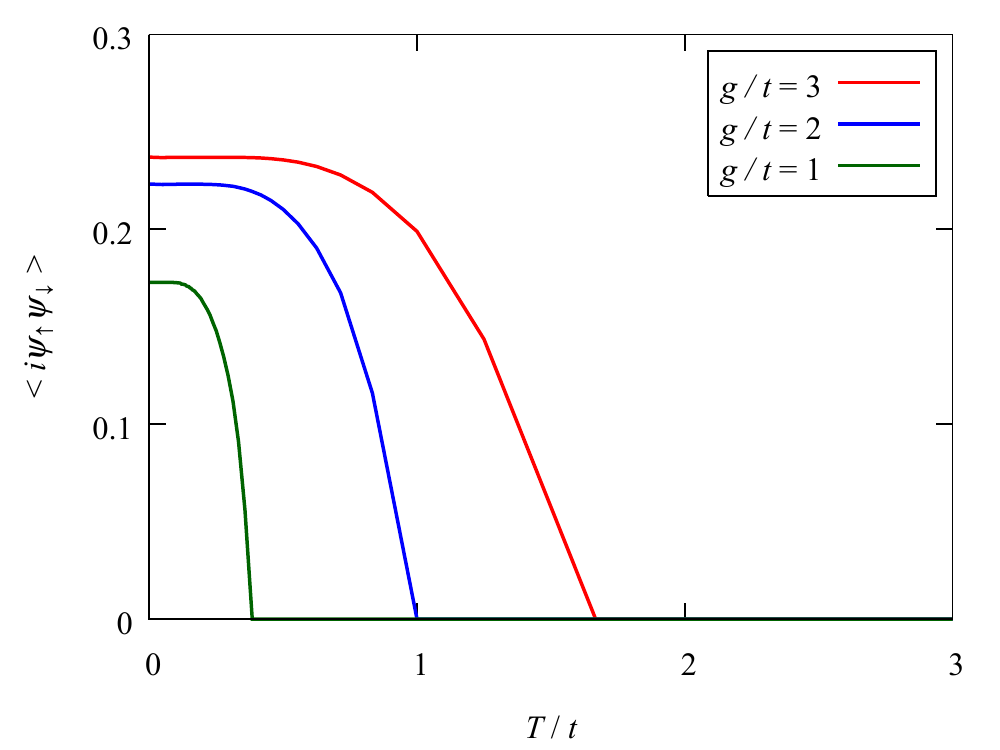}
\caption{
Mean-field results of the pair condensate $\langle i \psi_\uparrow \psi_\downarrow \rangle$ as a function of temperature $T$.
\label{figMF}
}
\end{figure}

Solving the gap equation \eqref{eqgap}, we obtain the pair condensate $\langle i \psi_\uparrow \psi_\downarrow \rangle$.
As mentioned in Sec.~\ref{secmodel}, the operator $i \psi_\uparrow \psi_\downarrow$ is time-reversal odd, so that the pair condensate $\langle i \psi_\uparrow \psi_\downarrow \rangle$ is an order parameter of spontaneous breaking of time-reversal symmetry.
Its temperature dependence is shown in Fig.~\ref{figMF}.
At high temperature, time-reversal symmetry is preserved and there appear gapless Majorana fermions.
At low temperature, time-reversal symmetry is broken and double Majorana fermions form a massive Dirac fermion.
The finite-size-scaling analysis shows that this is a second-order phase transition.
It corresponds to the phase transition to a density wave state in the Hamiltonian \eqref{eqHD2} \cite{PhysRevB.29.5253,PhysRevB.32.103}.

\section{Quantum Monte Carlo}
\label{secQMC}

Next, we study the phase structure in the quantum Monte Carlo method.
We introduce another type of the Hubbard-Stratonovich transformation to satisfy the Majorana positivity condition~\cite{2016CMaPh.346.1021J,2016PhRvL.116y0601W,Li:2016gte,Hayata:2017jdh}.
After the Hubbard-Stratonovich transformation, the path integral is
\begin{equation}
 Z = \int D\psi_\uparrow D\psi_\downarrow DA \ e^{ - S' - S_A } ,
\label{eqZ2}
\end{equation}
with
\begin{eqnarray}
&&
\begin{split}
S'
=& \int d\tau \sum_x \bigg[
\psi_\uparrow(x) \frac{\partial}{\partial\tau} \psi_\uparrow(x) + \psi_\downarrow(x)  \frac{\partial}{\partial\tau} \psi_\downarrow(x)
\\
&+ \sum_\mu [ \{ it + A_\mu(x) \} \psi_\uparrow(x) \psi_\uparrow(x+\hat{\mu})
\\
&+ \{ -it + A_\mu(x) \} \psi_\downarrow(x) \psi_\downarrow(x+\hat{\mu}) ] \bigg] ,
\end{split}
\\
&&
S_A = \int d\tau \sum_x \sum_\mu \frac{1}{2g} A_\mu^2 (x)
.
\end{eqnarray}

In the path-integral Monte Carlo simulation, imaginary time $\tau$ is discretized.
Naively, a temporal derivative is replaced by a central difference,
\begin{equation}
 \frac{\partial}{\partial\tau} \Psi(\tau) 
\to \hat{\partial}_\tau \Psi(\tau)
\equiv \frac{1}{2\delta\tau} \{ \Psi(\tau+\delta\tau) - \Psi(\tau-\delta\tau) \}
.
\end{equation}
However, the central difference has the doubling problem.
To understand this problem, let us consider the non-interacting propagator in momentum space
\begin{equation}
\frac{1}{M(\omega_n,p=0)} 
= \frac{1}{2} \frac{\delta\tau}{i\sin (\omega_n\delta\tau)}
,
\end{equation}
where $\omega_n$ is the fermionic Matsubara frequency.
The propagator has two poles at $\omega_n\delta \tau \to 0$ and $\pi$ (mod $2\pi$).
The pole at $\omega_n\delta \tau \to \pi$ is unphysical and thus called the doubler.
Although one well-known solution in non-relativistic theory is the use of a forward or backward difference, it does not solve the problem now.
In the case of the Majorana fermion, only the anti-symmetric part of the fermion matrix contributes to the path integral.
Because the anti-symmetric part of a forward or backward difference is a central difference, the calculation with a forward or backward difference is equivalent to that with a central difference.
To avoid the doubling problem, we discretize a temporal derivative as
\begin{equation}
\begin{split}
 \frac{\partial}{\partial\tau} \Psi(\tau) 
&\to \left(\hat{\partial}_\tau - \sigma_2 \frac{\hat{\Delta}_\tau}{2}\right) \Psi(\tau)
\\
&\equiv \frac{1}{2\delta\tau} \{ \Psi(\tau+\delta\tau) - \Psi(\tau-\delta\tau) \}
\\
&\quad - \sigma_2 \frac{1}{2\delta\tau} \{ \Psi(\tau+\delta\tau) + \Psi(\tau-\delta\tau) - 2 \Psi(\tau) \}
.
\label{eqdtlat}
\end{split}
\end{equation}
This is inspired by the Wilson fermion formalism in relativistic lattice field theory \cite{wilson1977quarks}.
Since the second term in Eq.~\eqref{eqdtlat} is a higher order of $\delta \tau$, it does not contribute to the $\delta \tau \to 0$ limit.
Now the non-interacting propagator becomes
\begin{equation}
\begin{split}
&\frac{1}{M(\omega_n,p=0)} 
\\
&= \frac{1}{2} \frac{\delta\tau}{i\sin (\omega_n\delta\tau) + \sigma_2 \{1-\cos (\omega_n\delta\tau)\}}
\\
&= - \frac{\delta\tau}{4} \frac{i\sin (\omega_n\delta\tau) - \sigma_2 \{1-\cos (\omega_n\delta\tau)\}}{1-\cos (\omega_n\delta\tau)}
.
\end{split}
\end{equation}
Only the physical pole at $\omega_n\delta \tau \to 0$ survives and the unphysical pole at $\omega_n\delta \tau \to \pi$ gets massive.
This resolves the doubling problem.
On behalf of it, time-reversal symmetry is explicitly broken at $\delta\tau \ne 0$.
This is similar to the explicit chiral symmetry breaking of the Wilson fermion formalism \cite{wilson1977quarks}.

After the Majorana fermions are integrated out, the path integral \eqref{eqZ2} becomes
\begin{equation}
 Z = \int DA \ \pf M \ e^{-S_A}
,
\label{eqZ3}
\end{equation}
where $\pf M $ is the Pfaffian of the anti-symmetric fermion matrix $M$.
The fermion matrix is given by
\begin{equation}
M =
\begin{pmatrix}
M_{\uparrow\uparrow} & M_{\uparrow\downarrow}
 \\
M_{\downarrow\uparrow} & M_{\downarrow\downarrow}
\end{pmatrix},
\end{equation}
with 
\begin{eqnarray}
&&
\begin{split}
 M_{\uparrow\uparrow} = 2 \hat{\partial}_\tau &+ \sum_\mu [ \{it+A_\mu(x)\}P_{+\mu} 
\\
&- \{it+A_\mu(x-\hat{\mu})\}P_{-\mu} ] ,
\end{split} 
\\
&&
\begin{split}
 M_{\downarrow\downarrow} = 2 \hat{\partial}_\tau &+ \sum_\mu [ \{-it+A_\mu(x)\}P_{+\mu}
\\
&- \{-it+A_\mu(x-\hat{\mu})\}P_{-\mu} ] ,
\end{split}
\\
&&
 M_{\uparrow\downarrow} = - M_{\downarrow\uparrow} = i \hat{\Delta}_\tau
.
\end{eqnarray}
This matrix satisfies the Majorana positivity condition~\cite{Hayata:2017jdh}.
The Pfaffian $\pf M$ is real and semi-positive.
Thus the path integral is free from the sign problem.
We remark here that the sub-lattice symmetry is not essential for the Majorana positivity of the model Hamiltonian~\eqref{eqH}.
It is still Majorana positive and does not suffer from the sign problem even for non-bipartite lattices such as a triangular lattice.
(On the other hand, for the case of repulsive interaction $g<0$, the auxiliary field $A_\mu$ is replaced by $i A_\mu$.
Then the fermion matrix no longer satisfies the Majorana positivity condition. 
The Pfaffian becomes indefinite and causes the sign problem.)

\begin{figure}[!t]
 \includegraphics[width=.48\textwidth]{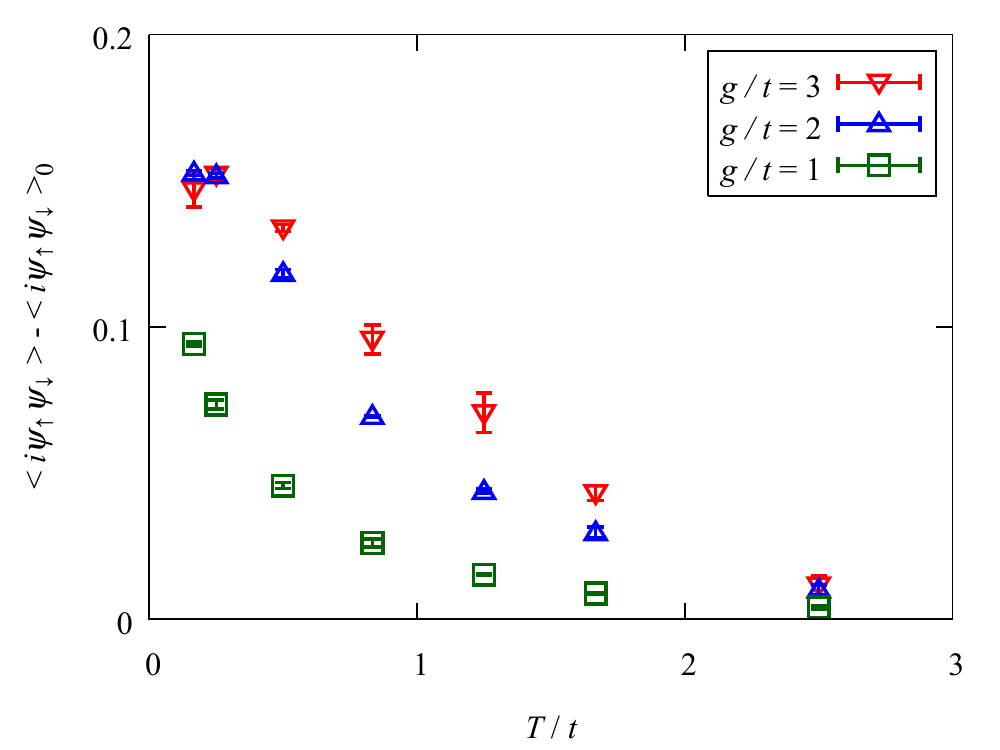}
\caption{
Monte Carlo data of the pair condensate $\langle i \psi_\uparrow \psi_\downarrow \rangle$ as a function of temperature $T$.
The non-interacting value $\langle i \psi_\uparrow \psi_\downarrow \rangle_0$ is subtracted from the Monte Carlo result $\langle i \psi_\uparrow \psi_\downarrow \rangle$. 
\label{figMC}
}
\end{figure}
\begin{figure}[t]
 \includegraphics[width=.48\textwidth]{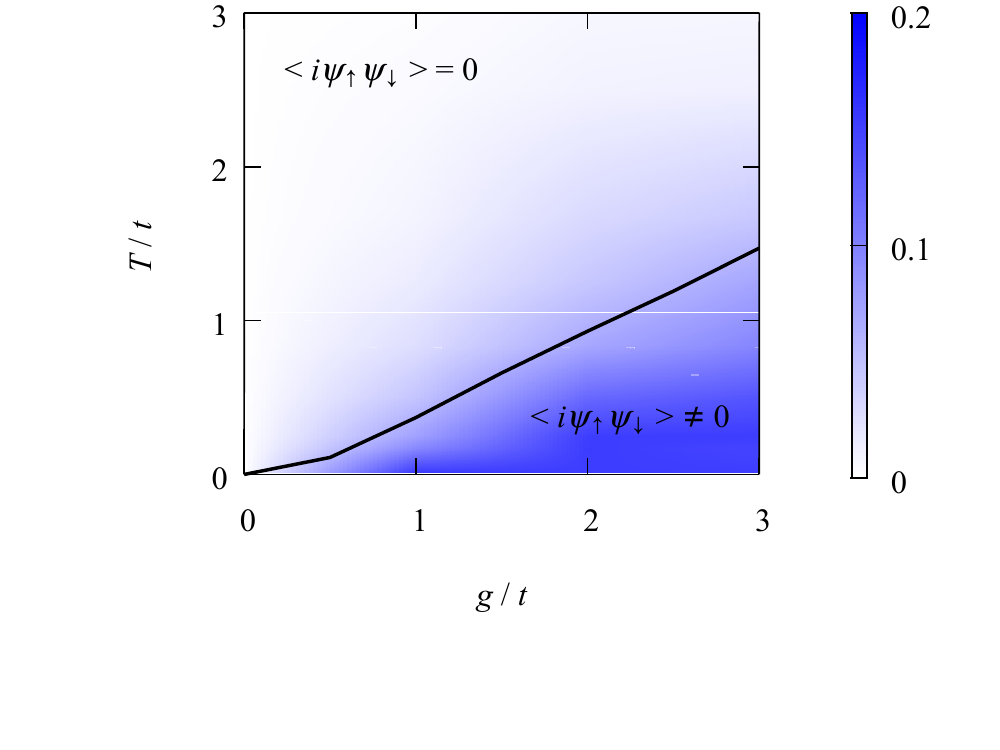}
\caption{
Phase diagram in the ($g,T$) plane.
The black solid line is the mean-field result of the phase transition line.
The blue gradation is the Monte Carlo data of the subtracted condensate $\langle i \psi_\uparrow \psi_\downarrow \rangle - \langle i \psi_\uparrow \psi_\downarrow \rangle_0$.
\label{figPhase}
}
\end{figure}

We performed the path-integral Monte Carlo simulation by the Hybrid Monte Carlo algorithm \cite{duane1987hybrid}.
For applying the Hybrid Monte Carlo algorithm, we transformed the path integral \eqref{eqZ3} to
\begin{equation}
\begin{split}
 Z  
&= \int DA \ (\det MM^\dagger)^{\frac{1}{4}} \ e^{-S_A}
\\
&= \int D\Phi^* D\Phi DA \ e^{ - \frac{1}{4} \Phi^\dagger (MM^\dagger)^{-1} \Phi - S_A }
.
\label{eqHMC}
\end{split}
\end{equation}
The complex scalar field $\Phi$ is called the pseudo-fermion field.
We note that the transformation \eqref{eqHMC} is invalid if the Pfaffian is indefinite.
We extrapolated the infinite-volume limit $V\to \infty$ from three lattice volumes $V=8^2$, $12^2$, and $16^2$.
We fixed the hopping parameter $t$ and the temporal discretization $\delta\tau$ at $t\delta\tau=0.1$.
Taking the continuum limit $\delta\tau \to 0$ is a future work.

The simulation results are shown in Fig.~\ref{figMC}.
The non-interacting value $\langle i \psi_\uparrow \psi_\downarrow \rangle_0$, which is induced by the explicit symmetry breaking \eqref{eqdtlat}, is subtracted from the Monte Carlo data $\langle i \psi_\uparrow \psi_\downarrow \rangle$. 
The results are qualitatively consistent with the mean-field results.
The symmetry is broken at low temperature and restored at high temperature.
However, the phase transition seems crossover, not second order.
This is the artifact of the explicit symmetry breaking \eqref{eqdtlat}.
To determine the order of the physical phase transition, we need to take the continuum limit $\delta\tau \to 0$ or formulate a discretization scheme to preserve the symmetry.

The phase diagram in the $(g,T)$ plane is shown in Fig.~\ref{figPhase}.
The second-order phase transition line in the mean-field approximation is also shown.
The Monte Carlo data and the mean-field results are qualitatively consistent.
Note that the temperatures of the path-integral Monte Carlo simulations are nonzero and the zero-temperature limit is obtained by the extrapolation of them.
The extrapolated data are shown at $T=0$ in Fig.~\ref{figPhase}.

\section{Summary} 
\label{sec:sum}
We have introduced the effective Majorana lattice model describing the low-energy physics of the vortex lattice in time-reversal-invariant topological superconductors.
We have studied spontaneous breaking of time-reversal symmetry at finite temperature.
To perform ab-initio analysis, we implemented the path-integral Monte Carlo simulation for Majorana fermion systems.
For attractive interaction, our model is free from the sign problem for any lattice structures including non-bipartite lattices in any dimensions, and can be evaluated by using the standard importance sampling method.
We have shown that the phase diagram of Majorana fermion systems can be investigated in the same way as that of complex fermion systems.

Our formulation is applicable to general Majorana fermion systems.
For example, it can be applied to variants of the Kitaev spin model~\cite{Kitaev20062,PhysRevLett.99.247203,PhysRevB.79.024426,PhysRevB.92.115137} by expressing them with Majorana fermions through the Jordan-Wigner transformation~\cite{PhysRevLett.98.087204,PhysRevB.76.193101,PhysRevLett.113.197205}.
It will be interesting to study the spin liquid state by using our method.
Another application is the Majorana fermion in particle physics, such as neutrinos, supersymmetry, and quantum vortices in color superconductors~\cite{Nishida:2010wr,Yasui:2010yw,Eto:2013hoa}.

\begin{acknowledgements}
T.~H.~was supported by JSPS Grant-in-Aid for Scientific Research (Grant No.~JP16J02240).
A.~Y.~was supported by JSPS KAKENHI (Grant No.~JP15K17624).   
The numerical simulations were carried out on SX-ACE in Osaka University.
\end{acknowledgements}

\bibliography{majorana}

\end{document}